\documentclass[cameraready]{Interspeech}
\makeatletter
\renewcommand{\author}[3][]{%
    \def\author@affiliation{}%
    \def\author@equalcontribution{}%
    \def\author@correspondingauthor{}%
    \def\author@orcid{}%
    \setkeys{authorkeys}{#1}%
    \ifx\author@orcid\@empty
        \edef\metadataOrcid{\metadataOrcid ; }%
        \def\orcid{}%
    \else
        \edef\metadataOrcid{\metadataOrcid \author@orcid; }%
        \def\orcid{\orcidURL{\author@orcid}}%
    \fi
    \ifx\author@affiliation\@empty
        \ifx\author@equalcontribution\@empty
            \ifx\author@correspondingauthor\@empty
                \protected@edef\authorlist{\authorlist\authorsep{}#2 #3\orcid}%
                \edef\metadataauthors{\metadataauthors #2 \{#3\}; }%
            \else
                \protected@edef\authorlist{\authorlist\authorsep{}#2 #3\orcid$^{\dagger}$}%
                \edef\metadataauthors{\metadataauthors #2 \{#3\}; }%
                \global\correspondingauthortrue
            \fi
        \else
            \ifx\author@correspondingauthor\@empty
                \protected@edef\authorlist{\authorlist\authorsep{}#2 #3\orcid$^{*}$}%
                \edef\metadataauthors{\metadataauthors #2 \{#3\}; }%
                \global\equalcontributiontrue
            \else
                \protected@edef\authorlist{\authorlist\authorsep{}#2 #3\orcid$^{*,\dagger}$}%
                \edef\metadataauthors{\metadataauthors #2 \{#3\}; }%
                \global\equalcontributiontrue
                \global\correspondingauthortrue
            \fi
        \fi
    \else
        \ifx\author@equalcontribution\@empty
            \ifx\author@correspondingauthor\@empty
                \protected@edef\authorlist{\authorlist\authorsep{}#2 #3\orcid$^{\author@affiliation}$}%
                \edef\metadataauthors{\metadataauthors #2 \{#3\} \{\author@affiliation\}; }%
            \else
                \protected@edef\authorlist{\authorlist\authorsep{}#2 #3\orcid$^{\author@affiliation,\dagger}$}%
                \edef\metadataauthors{\metadataauthors #2 \{#3\} \{\author@affiliation\}; }%
                \global\correspondingauthortrue
            \fi
        \else
            \ifx\author@correspondingauthor\@empty
                \protected@edef\authorlist{\authorlist\authorsep{}#2 #3\orcid$^{\author@affiliation,*}$}%
                \edef\metadataauthors{\metadataauthors #2 \{#3\} \{\author@affiliation\}; }%
                \global\equalcontributiontrue
            \else
                \protected@edef\authorlist{\authorlist\authorsep{}#2 #3\orcid$^{\author@affiliation,*,\dagger}$}%
                \edef\metadataauthors{\metadataauthors #2 \{#3\} \{\author@affiliation\}; }%
                \global\equalcontributiontrue
                \global\correspondingauthortrue
            \fi
        \fi
    \fi
    \renewcommand{\authorsep}{, }%
}

\renewcommand\maketitle{\par
    \begingroup
    \if@twocolumn
        \twocolumn[\@maketitle]
    \else \newpage
        \global\@topnum\z@ \@maketitle
    \fi\@thanks
    \endgroup
    \ifcameraready
        \ifequalcontribution
            \nolink{%
                \def\thefootnote{*}%
                \footnotetext{Equal Contribution}%
                \def\thefootnote{\arabic{footnote}}%
            }%
        \fi
        \ifcorrespondingauthor
            \nolink{%
                \def\thefootnote{$\dagger$}%
                \footnotetext{Corresponding Author}%
                \def\thefootnote{\arabic{footnote}}%
            }%
        \fi
    \fi
}
\makeatother
\title{Harf-Speech: A Clinically Aligned Framework for Arabic Phoneme-Level Speech Assessment}

\author[affiliation={1}, equalcontribution, correspondingauthor, orcid={0009-0007-1227-4024}]{Asif}{Azad}

\author[affiliation={1}, equalcontribution, orcid={0009-0000-1885-626X}]{MD Sadik Hossain}{Shanto}

\author[affiliation={1}, equalcontribution, orcid={0009-0003-3735-7046}]{Mohammad Sadat}{Hossain}

\author[affiliation={1}, orcid={0000-0002-3503-9138}]{Bdour}{Alwuqaysi}

\author[affiliation={1}, orcid={0000-0003-2734-3356}]{Sabri}{Boughorbel}

\author[affiliation={1}, orcid={0000-0002-2352-1701}]{Yahya}{Bokhari}

\author[affiliation={1}, orcid={0000-0002-5592-2462}]{Abdulrhman}{Aljouie}

\author[affiliation={2}]{Ayah Othman}{Sindi}

\author[affiliation={1, 3}, orcid={0000-0003-4781-4733}]{Ehsan}{Hoque}

\address{
    $^1$ Ministry of Defense, Saudi Arabia \\
    $^2$ Ability Center, Saudi Arabia \\
    $^3$ University of Rochester, USA 
}

\email{asifazad0178@gmail.com, shantosadikrglhs@gmail.com, sadat@cse.buet.ac.bd, bdour.alwuqaysi@hotmail.com, sabri.boughorbel@gmail.com, yahya.bokhari@gmail.com, aljouie@gmail.com, asindi@abilitycenter.sa, mehoque@cs.rochester.edu}


\usepackage{comment}
\usepackage{tabularx}
\usepackage{float}
\usepackage{fancyvrb}

\begin{document}

\maketitle


\begin{abstract}
    Automated phoneme-level pronunciation assessment is vital for scalable speech therapy and language learning, yet validated tools for Arabic remain scarce. We present Harf-Speech, a modular system scoring Arabic pronunciation at the phoneme level on a clinical scale. It combines an MSA phonetizer, a fine-tuned speech-to-phoneme model, Levenshtein alignment, and a blended scorer using longest common subsequence and edit-distance metrics. We fine-tune three ASR architectures on Arabic phoneme data and benchmark them with zero-shot multimodal models; the best, \textit{OmniASR-CTC-1B-v2}, achieves \textbf{8.92\%} phoneme error rate. Three certified speech-language pathologists independently scored 40 utterances for clinical validation. Harf-Speech attains a Pearson correlation of \textbf{0.791} and ICC(2,1) of \textbf{0.659} with mean expert scores, outperforming existing end-to-end assessment frameworks. These results show Harf-Speech yields clinically aligned, interpretable scores comparable to inter-rater expert agreement.
\end{abstract}

\section{Introduction}

Accurate pronunciation assessment is fundamental to speech therapy, assistive communication, and language learning. In speech-language pathology, evaluation at the phoneme level; analyzing substitutions, deletions, and distortions is essential for diagnosing articulation deficits and tracking progress. However, reliance on trained specialists limits scalability, motivating the need for reliable automated systems.

Arabic, spoken by over 400 million people and official in more than 20 countries, remains under-resourced in clinically grounded speech assessment technology. Modern Standard Arabic (MSA) poses particular challenges due to its rich consonantal inventory, emphatic and pharyngeal phonemes, and the functional role of short vowels and diacritics—all increasing phoneme-level sensitivity. While complete digital pronunciation assessment services such as Microsoft Azure Pronunciation Assessment exist, they operate as proprietary, one-size-fits-all systems that are neither localized for Arabic phonological characteristics nor evaluated against expert speech-language pathologist (SLP) judgments. Consequently, their clinical validity and suitability for Arabic speech therapy remain unclear.

We introduce \textbf{Harf-Speech}, a complete and clinically aligned framework for phoneme-level Arabic pronunciation assessment that integrates speech-to-phoneme modeling, reference phonetic generation, and automated scoring into a unified, interpretable system. As a core component, we fine-tune multiple automatic speech recognition (ASR) architectures with phoneme-level supervision and benchmark them alongside zero-shot multimodal models. Unlike proprietary approaches, Harf-Speech is built from open components and explicitly optimized for Arabic phonology. To ensure clinical relevance, three certified SLPs, each with over eight years of experience, independently scored utterances for inter-rater and system-expert alignment analysis.

Our contributions are threefold:
\begin{itemize}
    \item We present a clinically validated, complete framework for phoneme-level Arabic pronunciation assessment, addressing limitations of proprietary one-size-fits-all systems.
    \item We fine-tune and benchmark multiple ASR architectures for Arabic phoneme prediction, demonstrating substantial improvement over zero-shot multimodal and commercial baselines.
    \item We establish expert-aligned evaluation through direct comparison with certified SLP judgments, providing clinical grounding and reproducibility.
\end{itemize}

By demonstrating how open, localized modeling can outperform generic proprietary systems, this work advances scalable and clinically meaningful pronunciation assessment for Arabic and provides a replicable blueprint for other languages. 

\section{Related Work}

\begin{figure*}[t]
    \centering
    \includegraphics[width=0.75\linewidth]{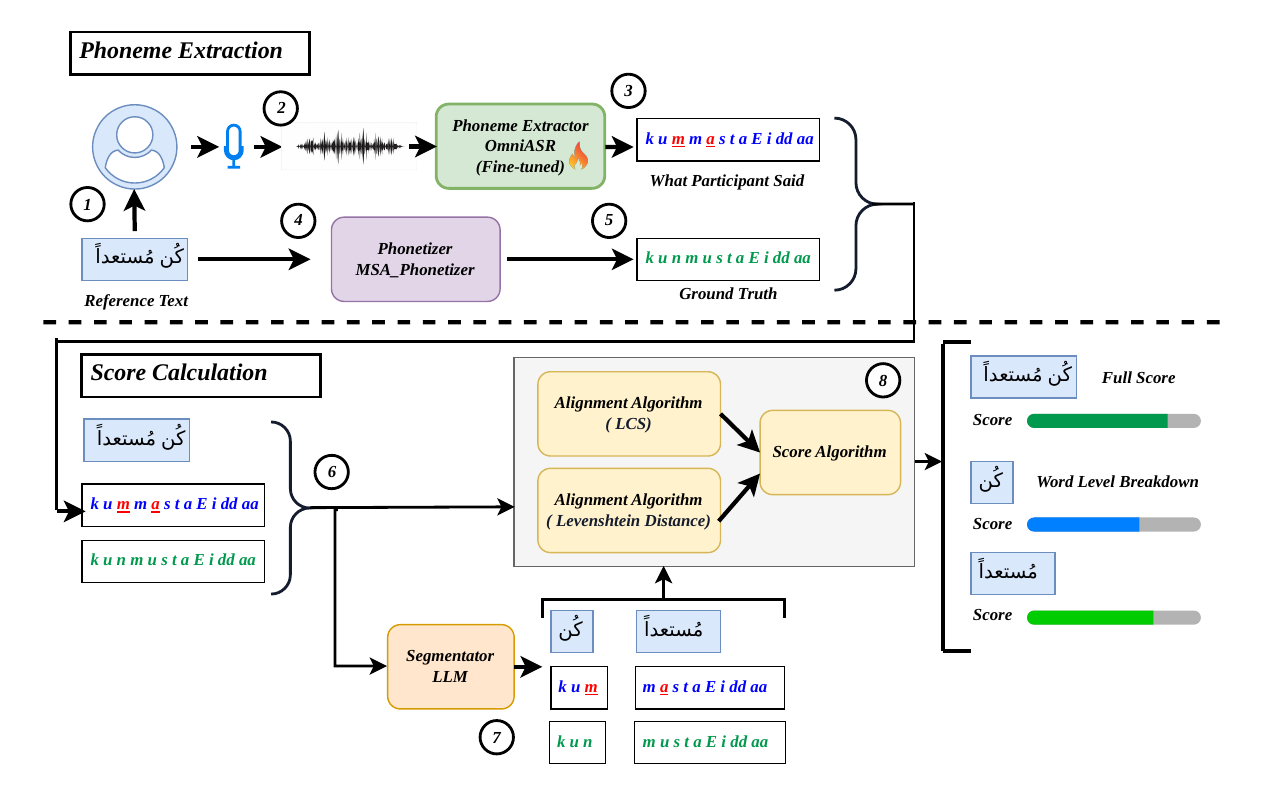}
    \caption{Overview of the Harf-Speech methodology. The example reference word shown in the figure translates to ``be prepared''}
    \label{fig:harf-overview}
\end{figure*}

Automated pronunciation assessment has evolved from Goodness-of-Pronunciation (GOP) scores derived from ASR likelihoods \cite{witt2000phone,kanters2009goodness} to deep learning approaches leveraging Connectionist Temporal Classification (CTC)-based architectures and self-supervised encoders such as wav2vec~2.0 \cite{baevski2020wav2vec}. While these advances have significantly improved phoneme recognition in high-resource languages, Arabic phoneme recognition remains comparatively underexplored. Prior Arabic studies have applied convolutional neural networks (CNNs), transformers, and transfer learning for mispronunciation detection \cite{alrashoudi2025improving,ccalik2024novel}, often focusing on Qur’anic recitation. However, challenges such as diacritic omission, rich phonemic contrasts, and limited standardized benchmarks restrict clinically aligned phoneme-level evaluation. Recent multimodal foundation models capable of processing audio input have also emerged, but their effectiveness for fine-grained Arabic phoneme transcription remains largely unexamined. Recently, QuranMB.v1 \cite{kheir2025towards} introduced the first publicly available benchmark for MSA phoneme-level mispronunciation detection, providing standardized evaluation protocols and baseline models that underscore the intrinsic difficulty of Arabic pronunciation assessment and the need for more advanced modeling approaches.

Phoneme recognition represents only one component of a complete digital pronunciation assessment framework. Complete systems of this type remain limited, with commercial platforms such as Microsoft Azure Cognitive Services offering proprietary, closed-box solutions without transparent phoneme-level modeling or clinical validation tailored to Arabic. Harf-Speech bridges these gaps by providing an open, modular, clinically validated phoneme-level scoring framework optimized for Arabic and readily extensible to other languages and future speech models.

\section{System Architecture}
\subsection{Overview}

Figure~\ref{fig:harf-overview} illustrates the overall architecture of Harf-Speech, an automated phoneme-level pronunciation assessment system designed to provide structured and interpretable feedback for Arabic speech production. At a high level, the system begins by presenting a reference sentence to the participant. The participant reads the sentence aloud, and the spoken response is recorded as an audio signal. The system then automatically processes both the reference text and the recorded audio, compares the expected and produced phoneme sequences, and generates a pronunciation score along with detailed word-level feedback.

From a system perspective, the architecture consists of four main stages: reference phoneme generation, speech-to-phoneme prediction, phoneme segmentation and alignment, and scoring algorithm. Each stage operates automatically without manual intervention, enabling scalable and consistent evaluation.

The design is modular and model-agnostic. While we have used our best performing fine-tuned model in this work for speech-to-phoneme prediction, alternative backbone models can be integrated without modifying the alignment and scoring framework. This flexibility ensures extensibility across datasets, dialects, and deployment settings.

\subsection{Reference Phoneme Generation}

Given a reference sentence, we generate its canonical phoneme sequence using an MSA-based phonetizer\footnote{\url{https://github.com/Iqra-Eval/MSA_phonetiser}}. The output is normalized into the Harf phoneme alphabet by removing positional suffixes and silence markers, resolving geminates, and remapping out-of-vocabulary symbols. This sequence serves as the expected ground-truth articulation.

\subsection{Speech-to-Phoneme Prediction}

The participant’s speech is converted directly into phoneme labels using a speech-to-phoneme model, enabling comparison at the minimal linguistic unit level. The framework is model-agnostic; however, off-the-shelf models underperform for Arabic phoneme recognition. We therefore fine-tuned multiple state-of-the-art architectures on Arabic phoneme data, with OmniASR-CTC-1B-v2 achieving the best performance.

\subsection{Segmentation and Alignment}

For word-level analysis, reference and predicted phoneme sequences are segmented into word-aligned groups using a large language model (LLM) conditioned on the text and phoneme sequences. Phoneme-level alignment is then computed via Levenshtein distance, yielding substitution, insertion, and deletion mappings for downstream scoring.

\subsection{Scoring Algorithm}

Given the aligned phoneme sequences, Harf-Speech computes a pronunciation score reflecting both sequential fidelity and edit-distance accuracy. Two complementary metrics are calculated per utterance:

\noindent\textbf{LCS Ratio.} The longest common subsequence (LCS) between reference and predicted phoneme sequences is divided by the reference length and scaled to $[0, 100]$. This captures preservation of phoneme order while tolerating isolated insertions.

\noindent\textbf{Pronunciation Score.} From the Levenshtein alignment, substitutions ($S$), deletions ($D$), and insertions ($I$) are obtained. Two sub-scores are defined:
\begin{equation}
    \text{Accuracy} = \max\left(0, \frac{N - S - D - I}{N}\right) \times 100
\end{equation}
\begin{equation}
    \text{Completeness} = \frac{N - D}{N} \times 100
\end{equation}
where $N$ denotes the number of reference phonemes. The pronunciation score is computed as:
\[
\text{PronScore} = 0.60 \times \text{Accuracy} + 0.40 \times \text{Completeness}.
\]

The final Harf-Speech score is:
\begin{equation}
    \text{Harf-Speech Score} = w_{\text{lcs}} \cdot \text{LCS Ratio} + w_{\text{pron}} \cdot \text{PronScore}
\end{equation}
with default empirical weights $w_{\text{lcs}} = 0.6$ and $w_{\text{pron}} = 0.4$, linearly mapped to a 0--5 clinical scale.

\section{Experimental Setup}

\subsection{Dataset}
For fine-tuning our phoneme-level ASR models, we primarily used the IqraEval dataset \cite{kheir2025towards}, which contains fully vowelized Modern Standard Arabic speech. We employed all three splits for training, including native and synthetic mispronounced samples. For benchmarking, we evaluated models on a randomly selected subset of 500 samples from the validation split. This subset evaluation was necessary due to the very high inference time observed for the Gemini-3-pro model and, to a lesser extent, Gemini-3-flash, which made full validation-set inference computationally impractical.

The training data combines three sources to ensure robustness and coverage of realistic pronunciation variations: (1) Native speech from fluent MSA speakers, treated as “golden” pseudo-labeled phonemes, providing accurate canonical references; (2) Synthetic mispronunciations generated via text-to-speech (TTS) systems, where phoneme-level errors were systematically introduced using a confusion matrix to simulate common pronunciation mistakes; and (3) Real mispronunciations recorded from human speakers, capturing authentic deviations and prosodic variability. 

For clinical validation, we curated 40 speech samples from diverse sources and had them independently scored by three certified native Arabic-speaking speech-language pathologists (SLPs), each with 8–10 years of clinical experience in Arabic speech and language assessment. Forty samples were selected to ensure sufficient variability for reliability analysis while maintaining feasible expert annotation effort, and three raters were used as the minimum required to compute stable inter-rater agreement statistics without redundancy. Each SLP rated the samples on a 0–5 pronunciation scale. This expert-rated subset was then used to evaluate the alignment between automated phoneme-level scoring and clinician judgments. Details of the sample curation are provided in \textit{Supplementary Material}.

\begin{table}[t]
    \centering
    \small
    \setlength{\tabcolsep}{3pt}
    \caption{Phoneme error rate (PER) and real-time factor (RTF; inference time divided by audio duration) comparison across models. ZS = zero-shot; FT = fine-tuned.}
    \label{tab:per_results}
    \begin{tabularx}{\linewidth}{X c c >{\raggedleft\arraybackslash}p{1.0cm} >{\raggedleft\arraybackslash}p{1.0cm}}
        \toprule
        Model             & \shortstack{Params} & Type        & PER             & RTF            \\
        \midrule
        Gemini-3-flash    & unknown                  & \textsc{ZS} & 17.31\%         & 0.394          \\
        Gemini-3-pro      & unknown                  & \textsc{ZS} & 15.07\%         & 10.748         \\
        Qwen3-ASR-1.7B    & 1.7B                & \textsc{FT} & 16.79\%         & 0.130          \\
        Wav2Vec2-LV60-CV  & 317M                & \textsc{FT} & 13.58\%         & 0.009          \\
        OmniASR-CTC-1B-v2 & 1B                  & \textsc{FT} & \textbf{8.92\%} & \textbf{0.004} \\
        \bottomrule
    \end{tabularx}
\end{table}

\subsection{Fine-tuning Details}

We fine-tuned three state-of-the-art ASR models on the phoneme-level Arabic dataset (Section 3.1): 
\begin{itemize}
    \item \textit{Wav2Vec2-LV-60-ESpeak-CV-FT} \cite{xu2021simple}
    \item \textit{Qwen3-ASR-1.7B} \cite{Qwen3-ASR}
    \item \textit{OmniASR-CTC-1B-v2} \cite{omnilingualasr2025}
\end{itemize}
adapting them for robust phoneme prediction across canonical, synthetic, and real mispronounced speech.

All models were trained using mixed precision (FP16/BF16), gradient accumulation (×4), and linear learning rate scheduling. For Wav2Vec2, we used an effective batch size of 64, learning rate 3e-5, weight decay 0.01, and trained for 10 epochs with PER-based model selection. OmniASR-CTC-1B-v2 was trained for 35k steps with validation every 1k steps, while Qwen3-ASR followed a comparable setup.

% \subsection{Finetuning Details}
% We fine-tuned three state-of-the-art ASR models on the phoneme-level Arabic dataset described in Section 3.1:
% \begin{itemize}
%     \item \textit{Wav2Vec2-LV-60-ESpeak-CV-FT} \cite{xu2021simple}
%     \item \textit{Qwen3-ASR-1.7B} \cite{Qwen3-ASR}
%     \item \textit{OmniASR-CTC-1B-v2} \cite{omnilingualasr2025}
% \end{itemize}
% Fine-tuning aimed to adapt these models for robust phoneme-level prediction across canonical, synthetic, and real mispronounced speech.

% For Wav2Vec2, fine-tuning used a per-device batch size of 16 with gradient accumulation of 4 (effective batch size 64), mixed-precision training (FP16), and gradient checkpointing to reduce memory usage. The learning rate was set to 3e-5 with a weight decay of 0.01, and the models were trained for 10 epochs with 3,200 warmup steps. Evaluation and checkpointing were performed every 2,000 steps, with the best model selected based on phoneme error rate (PER).

% Qwen3-ASR-1.7B and OmniASR-CTC-1B-v2 were fine-tuned using similar strategies, with mixed-precision (BF16 for Qwen3, FP16 for OmniASR), gradient accumulation of 4, and linear learning rate schedules. Training steps were set to 35,000 for OmniASR-CTC-1B-v2, with validation and checkpointing every 1,000 steps. 

\section{Results \& Analysis}

\begin{figure*}[t]
    \centering
    \includegraphics[width=0.9\linewidth]{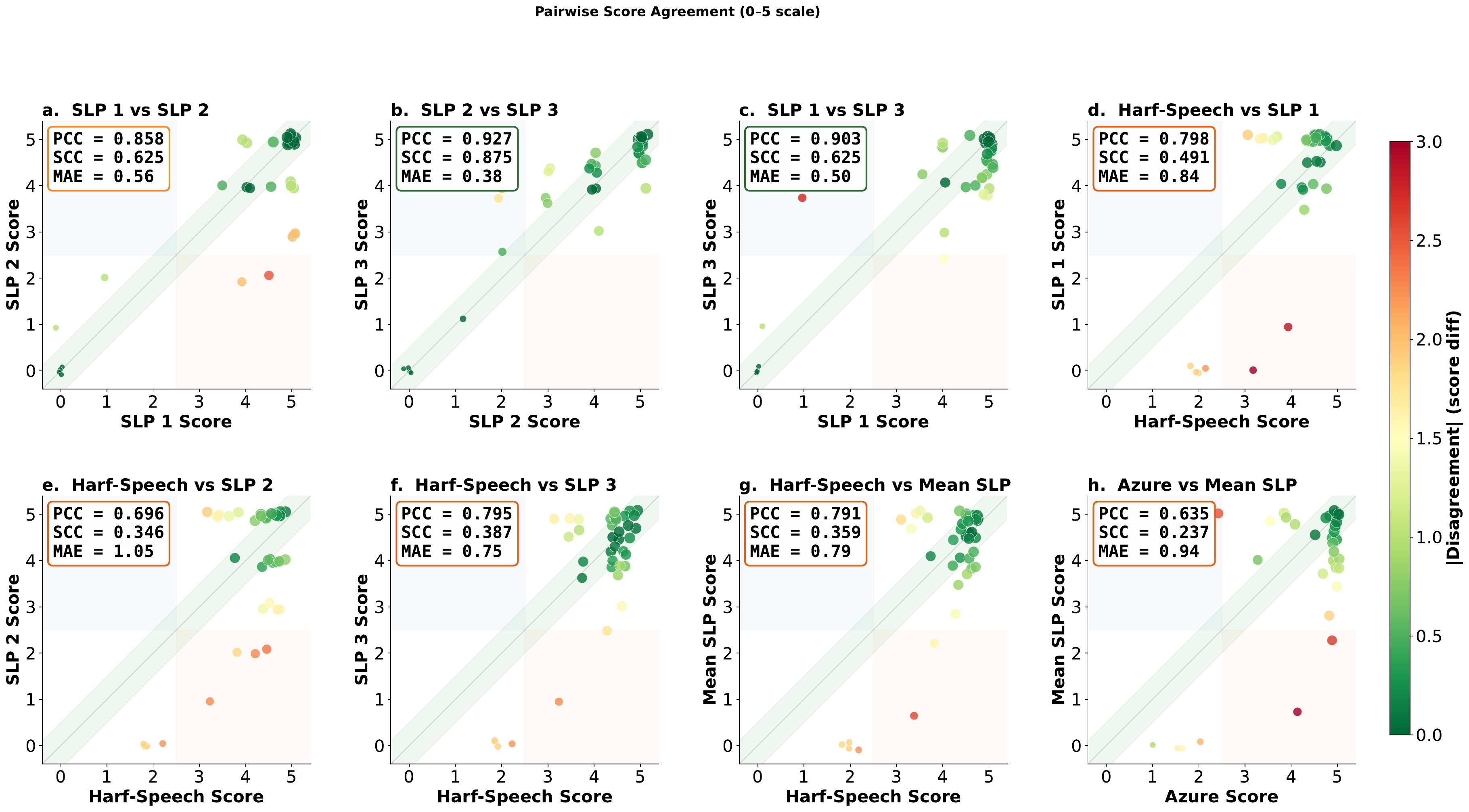}
    \caption{Pairwise scatter plots on the 0--5 clinical scale. Points are colored by absolute disagreement (green = close, red = far). The shaded band marks $\pm$0.5 tolerance around perfect agreement.}
    \label{fig:scatter_8panel}
\end{figure*}

\subsection{Phoneme Recognition}
Table~\ref{tab:per_results} summarizes the phoneme error rate (PER) and real-time factor (RTF) across all evaluated models; metric definitions are provided in \textit{Supplementary Material}. For zero-shot multimodal models (Gemini and Qwen-Omni), we used a standardized SLP-style system prompt to elicit phoneme sequences; the full system prompt and vocabulary reference are provided in \textit{Supplementary Material}. Fine-tuned models consistently outperform zero-shot multimodal models. OmniASR-CTC-1B-v2 achieves the lowest PER of 8.92\% while also being the fastest (RTF 0.004), making it the clear choice for the Harf-Speech backbone. Among zero-shot models, Gemini-3-pro-preview reaches 15.07\% PER but with a prohibitively high RTF of 10.75, while Qwen2.5-Omni-7B failed to produce usable phoneme outputs for Arabic. Wav2Vec2 attains 13.58\% PER with the second-lowest RTF, offering a lightweight alternative. These results confirm that task-specific fine-tuning remains essential for Arabic phoneme-level modeling.

\subsection{Inter-Rater Agreement \& Clinical Alignment}

Table~\ref{tab:full_agreement} reports pairwise agreement among the three SLPs and between each automated system and the mean expert score (see \textit{Supplementary Material} for metric definitions). The metrics include Pearson correlation coefficient (PCC), Spearman correlation coefficient (SCC), intraclass correlation coefficient (ICC), mean absolute error (MAE), root mean squared error (RMSE), exact agreement, and ±1 agreement. Inter-SLP agreement is strong (PCC 0.858--0.927, ICC $\geq$ 0.846), with $\pm$1 agreement rates of 84.6--94.9\%, confirming that expert phoneme-level scoring is highly reproducible and serves as a reliable reference.

Harf-Speech achieves a PCC of 0.791 and ICC(2,1) of 0.659 against the mean SLP score, with a $\pm$1 agreement of 76.9\%. Crucially, this performance is consistent across individual raters as well: Harf-Speech reaches PCC 0.798 against SLP~1 and 0.795 against SLP~3, approaching the lower end of the inter-SLP range (Figure ~\ref{fig:scatter_8panel}). In contrast, Azure Pronunciation Assessment achieves substantially lower correlation with every SLP (PCC 0.532--0.672) and higher error (MAE 0.82--1.14, RMSE 1.20--1.51). Against Mean SLP, Harf-Speech outperforms Azure by +0.156 in PCC, +0.066 in ICC, and reduces MAE by 16\% (0.79 vs.\ 0.94).

\begin{table}[ht!]
    \centering
    \small
    \setlength{\tabcolsep}{2pt}
    \caption{Full pairwise agreement metrics on the 0--5 clinical scale, grouped by SLP target. Bold values indicate the better result between Harf-Speech and Azure within each group.}
    \label{tab:full_agreement}
    \begin{tabularx}{\linewidth}{X c c c c c c c}
        \toprule
        Pair           & PCC           & SCC           & \shortstack{ICC                                                                 \\(2,1)} & MAE & RMSE & \shortstack{Exact\\(\%)} & \shortstack{$\pm$1\\(\%)} \\
        \midrule
        \multicolumn{8}{l}{\textit{Inter-SLP}}                                                                                           \\
        SLP 1 vs SLP 2 & .858          & .625          & .846            & 0.56          & 0.95          & 61.5          & 84.6          \\
        SLP 2 vs SLP 3 & .927          & .875          & .922            & 0.39          & 0.64          & 69.2          & 94.9          \\
        SLP 1 vs SLP 3 & .903          & .625          & .896            & 0.50          & 0.75          & 53.8          & 94.9          \\
        \midrule
        \multicolumn{8}{l}{\textit{vs.\ SLP 1}}                                                                                          \\
        Harf-Speech    & \textbf{.798} & \textbf{.491} & \textbf{.640}   & 0.84          & \textbf{1.15} & 43.6          & 76.9          \\
        Azure          & .634          & .445          & .590            & \textbf{0.82} & 1.34          & \textbf{48.7} & \textbf{79.5} \\
        \midrule
        \multicolumn{8}{l}{\textit{vs.\ SLP 2}}                                                                                          \\
        Harf-Speech    & \textbf{.696} & \textbf{.346} & \textbf{.555}   & \textbf{1.05} & \textbf{1.26} & 28.2          & 64.1          \\
        Azure          & .532          & .218          & .477            & 1.14          & 1.51          & \textbf{30.8} & 64.1          \\
        \midrule
        \multicolumn{8}{l}{\textit{vs.\ SLP 3}}                                                                                          \\
        Harf-Speech    & \textbf{.795} & \textbf{.387} & \textbf{.672}   & \textbf{0.75} & \textbf{1.02} & \textbf{35.9} & 76.9          \\
        Azure          & .672          & .312          & .631            & 0.90          & 1.20          & 30.8          & 76.9          \\
        \midrule
        \multicolumn{8}{l}{\textit{vs.\ Mean SLP}}                                                                                       \\
        Harf-Speech    & \textbf{.791} & \textbf{.359} & \textbf{.659}   & \textbf{0.79} & \textbf{1.05} & \textbf{35.9} & 76.9          \\
        Azure          & .635          & .237          & .593            & 0.94          & 1.28          & 33.3          & 76.9          \\
        \bottomrule
    \end{tabularx}
\end{table}

The scatter plots in Figure~\ref{fig:scatter_8panel} visualize per-sample agreement on the 0--5 clinical scale. In the inter-SLP panels~(a--c), points cluster tightly along the diagonal with predominantly green coloring, reflecting low disagreement. The Harf-Speech panels~(d--g) show a similar pattern: scores distribute closely around the identity line for all three individual SLPs and their mean, with the majority of samples falling within the $\pm$0.5 tolerance band. In contrast, the Azure panel~(h) exhibits noticeably wider scatter and more orange/red points, indicating larger per-sample deviations. These results demonstrate that Harf-Speech produces scores that are well-aligned with expert clinical judgments and substantially more reliable than a proprietary baseline.

\section{Conclusion}

We presented Harf-Speech, a complete automated framework for phoneme-level Arabic speech assessment validated against expert clinical judgments. By fine-tuning modern ASR architectures for speech-to-phoneme modeling, the system achieves strong clinical alignment with a \textbf{0.791} Pearson correlation to speech-language pathologist ratings. Its modular and open design enables straightforward integration of future state-of-the-art ASR models and adaptation to other languages beyond Arabic. By demonstrating how localized, clinically grounded modeling can outperform generic systems, Harf-Speech contributes toward more accessible, scalable, and reproducible pronunciation assessment technologies. The study was carried out under IRB approval from the Ministry of Defense, Saudi Arabia, and as part of the Digital Harf \cite{digital-harf} platform's SFDA-approved clinical trial.

\section{Acknowledgments}
The authors would like to thank Laila Shehab Salamah and Renad Sayegh, both certified Speech-Language Pathologists from King Fahad Armed Forces Hospital, for their invaluable clinical expertise and contributions to the validation of this study. Special thanks are also extended to Sama Almuraykhi and Sara Alghamdi from the Ministry of Defense Digital Transformation, Saudi Arabia. 

A special acknowledgement to Dr. Altom Saleh, Abdulaziz Alnujaidi, Dr. Wafa Alatar, Razan Albassam, and Sadeem Babnji, speech therapists from the Mohammed Bin Salman Autism Program, for their contributions and support.

Finally, we express our deep gratitude to Dr. Waleed Alhazzani and Dr. Hadeel Abdulmohsen AlKhamees for their strategic support throughout this research.

\section{Generative AI Use Disclosure}

The authors confirm that no generative AI tools were used to create substantive content in this manuscript. Generative AI was only used for language polishing, editing, and minor stylistic improvements to ensure clarity and readability. All (co-)authors take full responsibility for the scientific content, analysis, and conclusions presented in this work.    

% Prevent an empty thebibliography (which triggers "missing \item") when there are
% no \cite commands yet.
\nocite{*}

\bibliographystyle{IEEEtran}
\bibliography{mybib}

\end{document}